\documentclass[aps,prl,twocolumn,superscriptaddress,longbibliography]{revtex4-1}
\pdfoutput=1

\usepackage{graphicx} %zum Einbinden von Graphiken
\usepackage{amssymb}
\usepackage{amsmath}
\usepackage{mathtools}
\usepackage{ifthen}
\usepackage{braket}

\usepackage{footnote}

\usepackage{color}
\usepackage[colorlinks,bookmarks=false,citecolor=darkblue,linkcolor=red,urlcolor=blue]{hyperref}
\definecolor{darkred}{rgb}{0.7,0.0,0.0}
\definecolor{darkblue}{rgb}{0,0.02,0.45}
\definecolor{grey}{rgb}{0.5,0.5,0.5}

\usepackage[capitalise]{cleveref}

\newcommand{\GammaMag}[1][]{\ensuremath{\Gamma_{{B#1}}}}

\newcommand{\aRuCl}{$\alpha$-RuCl$_3$}
\newcommand{\Bc}[1][]{\ensuremath{B_c^{\mathrm{AF#1}}}}

\newcommand{\bdirection}{[010]}
\newcommand{\adirection}{[110]}

\newcommand{\Ctot}{\ensuremath{C_{\mathrm{tot}}}}

\newcommand{\Csample}{\ensuremath{C_{\mathrm{Sa}}}}
\newcommand{\Ccell}{\ensuremath{C_{\mathrm{Cell}}}}

\newcommand{\temp}[1][]{\ensuremath{T_{\mathrm{#1}}(t)}}

\newcommand{\rucl}{$\alpha$-RuCl$_3$}

\begin{document}

\title{Angle-dependent thermodynamics of $\alpha$-RuCl$_3$}

\author{S. Bachus}
\email[]{sebastian.bachus@physik.uni-augsburg.de}
\affiliation{Experimental Physics VI, Center for Electronic Correlations and Magnetism, University of Augsburg, 86159 Augsburg, Germany}

\author{D.~A.~S.~Kaib}
\email[]{kaib@itp.uni-frankfurt.de}
\affiliation{Institute of Theoretical Physics, Goethe University Frankfurt, 60438 Frankfurt am Main, Germany}

\author{A. Jesche}
\affiliation{Experimental Physics VI, Center for Electronic Correlations and Magnetism, University of Augsburg, 86159 Augsburg, Germany}

\author{V. Tsurkan}
\affiliation{Experimental Physics V, Center for Electronic Correlations and Magnetism, University of Augsburg, 86159 Augsburg, Germany}
\affiliation{Institute of Applied Physics, Chisinau MD-2028, Republic of Moldova}

\author{A.~Loidl}
\affiliation{Experimental Physics V, Center for Electronic Correlations and Magnetism, University of Augsburg, 86159 Augsburg, Germany}

\author{S.~M.~Winter}
\email[]{winters@wfu.edu}
\affiliation{Department of Physics and Center for Functional Materials, Wake Forest University, Winston-Salem, North Carolina 27109, United States}

\author{A. A. Tsirlin}
\email[]{altsirlin@gmail.com}
\affiliation{Experimental Physics VI, Center for Electronic Correlations and Magnetism, University of Augsburg, 86159 Augsburg, Germany}

\author{R.~Valent\'\i}
\email[]{valenti@itp.uni-frankfurt.de}
\affiliation{Institute of Theoretical Physics,
Goethe University Frankfurt, 60438 Frankfurt am Main, Germany}

\author{P. Gegenwart}
\email[]{philipp.gegenwart@physik.uni-augsburg.de}
\affiliation{Experimental Physics VI, Center for Electronic Correlations and Magnetism, University of Augsburg, 86159 Augsburg, Germany}

%\date{\today}

\begin{abstract}
Thermodynamics of the Kitaev honeycomb magnet $\alpha$-RuCl$_3$ is studied for different directions of in-plane magnetic field using measurements of the magnetic Gr\"uneisen parameter \GammaMag{} and specific heat $C$. We identify two critical fields \Bc[1] and \Bc[2] corresponding, respectively, to a transition between two magnetically ordered states and the loss of magnetic order toward a quantum paramagnetic state. The \Bc[2] phase boundary reveals a narrow region of magnetic fields where inverse melting of the ordered phase may occur. 
%reentrance with temperature. %reentrance of the ordered phase with increasing temperature. %is non-monotonic, suggesting abundance of quantum fluctuations in the vicinity of this transition. 
%\dkc{this could be toned down if we are not so sure.  ``reveals'' could be ``suggests''} 
	No additional transitions are detected above \Bc[2] for any direction of the in-plane field, although a shoulder anomaly in \GammaMag{} is observed systematically at $8-10$\,T.
Large field-induced entropy effects imply additional low-energy excitations at low fields and/or strongly field-dependent phonon entropies. 
 Our results establish universal features of $\alpha$-RuCl$_3$ in high magnetic fields and challenge the presence of a field-induced Kitaev spin liquid in this material.  %exclude \dk{Kitaev} spin liquid as a separate magnetic phase, and shed light on the anisotropy of the excitation gap as seen from the field dependence of \GammaMag{} and magnetic entropy. \dk{We do find more entropy along b than a, which fits tanakas scenario qualitatively (although it is weird that in his scenario the phase gets more stable with more magnetic field). However, overall entropy changes are too large to be from magnetic degrees of freedom only, and entropy seems to be saturated both for b and a. We therefore propose an alternative picture where 1) phonon entropy is strongly magnetic field dependent due to strong magnetoelastic coupling 2) additional low-energy modes, that are already present at zero field, that seem to be independent/additional to the bulk description of rucl (zigzags then polarized)}
\end{abstract}

\maketitle


%------------------------------------------------------------------------------------------------------
%Exploring new phenomena is one of the essential purposes in research. Since decades, the quest for Quantum Spin Liquids (QSL) 
The Kitaev honeycomb model offers a possible practical route toward a quantum spin liquid with exotic fractionalized excitations and potential applications in topological quantum computing~\cite{Kitaev,Hermanns2018}. The $4d$ layered honeycomb material $\alpha$-RuCl$_3$ is one of the best experimental realization of this model available so far~\cite{Rau2016,Winter2017,Takagi2019}. 
 However, it develops long-range magnetic order in zero magnetic field and may %\dk{or may not}
 only be proximate to the elusive Kitaev spin liquid (KSL)~\cite{Banerjee,Banerjee2017}. Tuning $\alpha$-RuCl$_3$ by pressure~\cite{Bastien2018,Biesner2018} or chemical substitution~\cite{lampenkelley2017} is hampered by unwanted effects of dimerization and magnetic dilution, respectively. On the other hand, magnetic fields applied within the honeycomb ($ab$) plane can act as a very clean tuning mechanism that eliminates magnetic order above $B_c=7.0-7.5$\,T~\cite{Sears2017,Baek2017,Banerjee2018}.

%\dk{$\phi=0$ is defined as $B\parallel b$ right now. We could change it to $\phi=0$ being $B\parallel a$. Then we would have the same definition as Tanaka.} 
The physics within this region without magnetic order are subject of significant debate. On the
one hand, a half-integer plateau in the thermal Hall conductivity (THC) as
expected for the KSL phase has been reported \cite{Kasahara2018}. However,  
%spectra seen in spectroscopic measurements in line with a topologically-trivial, partially-polarized phase \cite{Wulferding2020,Sahasrabudhe2020,Ponomaryov2020}. 
spectroscopic measurements find excitations that are expected for the topologically trivial, partially-polarized phase \cite{Sahasrabudhe2020,Ponomaryov2020,maksimov2020rethinking}. %\dk{Not sure if we can cite Ref.~\cite{Wulferding2020} here. They say they see Majoranas and Anyons in their data... RV: I agree, probably we either rephrase or take the citation out} 
To scrutinize this, recent studies %focused on / considered
%Recent studies structinized this by considering
 %the in-plane angle $\phi$ (see Fig.~X\dk{need figure with $\phi$}) of the magnetic field:
 compared %the effects of different in-plane magnetic field angles $\phi$ (see Fig.~X\dk{need figure with $\phi$})
 the effects from rotating the magnetic field within the plane 
  against the expectations for the KSL: 
 %Within the pure Kitaev model with small magnetic fields
% According to Kitaev's perturbation theory \cite{Kitaev}, 
%the KSL would be gapless for fields parallel to a bond ($\phi=0^\circ + n60^\circ,\, n\in \mathbb Z $), but gains a gap and non-Abelian topological order for other in-plane fields. While 
%
%In the Kitaev model, 
%\emph{most} in-plane fields angles $\phi$ would induce an excitation gap and %non-Abelian topological order
%, except for fields parallel to a bond ($\phi=0^\circ + n60^\circ,\, n\in \mathbb Z $), where the KSL would be gapless. 
%a non-zero Chern number that leads to the quantized THC with $\phi$-dependent sign. Yet for fields parallel to a bond ($\phi=0^\circ + n60^\circ,\, n\in \mathbb Z $), the KSL would be gapless and show no THC. 
For the Kitaev model under \emph{most} in-plane field angles $\phi$, the KSL would be gapped and show a quantized THC with $\phi$-dependent sign. Yet for fields parallel to a bond ($\phi=%0^\circ +
 n\cdot 60^\circ,\, n\in \mathbb Z $  \footnote{Throughout the manuscript, we discuss the Kitaev model and \rucl\ in the framework of $C_6$ symmetry within the honeycomb lattice, implying that field angles are equivalent up to rotations of $60^\circ$. This symmetry is likely only approximately present for \rucl\ in its $C2/m$ structure.}), the KSL would be gapless and show no THC. 
Experimentally, the sign structure of the THC was indeed observed in \rucl\ \cite{Yokoi2020}. 
However, for in-plane fields, the same sign structure is expected for the partially-polarized phase, suggesting that the dependence of the THC with $\phi$ alone at the accessed temperatures cannot distinguish between the scenarios \cite{chern2020SignStructureThermal}. 
%However, it was also argued, that a similar structure of the THC, aside from its precise $$
%However, it was later noted that the same sign structure is also expected for the partially-polarized phase, implying that the sign structure alone cannot distinguish between the scenarios \cite{chern2020SignStructureThermal}. 
%However, it was later shown that a compatible sign %the same sign 
%structure is also expected in the partially-polarized phase% in more generic models
%, suggesting that the sign structure of THC with $\phi$ alone cannot distinguish between the scenarios \cite{chern2020SignStructureThermal}. \dk{Actually, The Yokoyoi paper also investigates and argues with out-of-plane signs, which the Chern paper doesnt touch, so I toned it down to suggest} 
The gaplessness of the field-induced state for distinct $\phi$ (as predicted for the KSL) however would be a property clearly incompatible with the partially-polarized phase. 
Accordingly, the $\phi$-dependent gap was recently investigated by Tanaka {\it et al.}~\cite{Tanaka2020}. From analysis of their low-temperature specific heat measurements, a gapless behavior for $\phi=%0^\circ +
 n\cdot 60^\circ$ and gapped behavior for $\phi\neq%0^\circ +
 n\cdot 60^\circ$, as expected for the KSL, was reported~\cite{Tanaka2020}. 
Another pertinent question is the occurrence of a phase transition upon entering and leaving this putative QSL as a function of field.
%While several possible signatures of a field-induced KSL were reported in the recent literature, %~\cite{Kasahara2018,Yokoi2020,Tanaka2020}, %the formation of the KSL as a stable \textit{phase} of $\alpha$-RuCl$_3$ requires that phase transitions occur upon entering and leaving the topological KSL regime as a function of field
%the formation of the KSL as a gapped topological phase of $\alpha$-RuCl$_3$
%%\dkc{Actually we also expect strong signatures like a phase transition for the gapless KSL... But it is not strictly required.}
% requires that phase transitions occur upon entering and leaving the KSL as a function of field. 
No conspicuous signatures of such transitions %\dk{have to make clear that we only need to find one transition... not two.. Bc2 could be the left one} 
are found in recent studies using various thermodynamic probes \cite{Schoeneman2020,Bachus2020}. %\dk{Gass says they see a first-order transition at 11T. RV: Correct, we should be careful with this citation}
%various thermodynamic probes, for example in the recent studies on magnetostriction for $\phi=0^\circ$\cite{Gass2020} \cite{Schoeneman2020,Bachus2020}. 
  Here, we employ the magnetic Gr\"uneisen parameter ($\GammaMag$) and specific heat ($C$) to map out the field-angle, field-strength and temperature-dependent phase diagram of $\alpha$-RuCl$_3$.  
%and demonstrate the absence of such transitions within the quantum disordered region 
%\dkc{added this because we do find of course $Bc^{AF2}$, which \textit{could} be the entering point of the KSL. We just dont find the exit point. }
%These demonstrate the absence of any further transitions within the quantum disordered region \textcolor{blue}{up to $14\,\text T$} for \emph{any} direction of  the in-plane magnetic field. 
These demonstrate the absence of further phase transitions  beyond the magnetically ordered region for \emph{any} in-plane angle of the magnetic field.  %\dkc{One could also add here the reentrant phase behavior, but I think it doesnt fit the flow. in Abstract and conclusions then. }
We furthermore uncover a finite region where inverse melting of the antiferromagnetic phase may occur and track shoulder-anomalies in $\GammaMag{}$ within the quantum paramagnetic region. 
%Our measured phase boundaries of the antiferromagnetic phases reveal a
%This establishes the relevant phase diagram as a function of these parameters. 
In addition, the combination of $\GammaMag{}$ and $C$ allows to directly resolve entropy-differences between states at different field strengths and angles. 
These reveal an unexpectedly large loss of entropy from the zero-field gapped ordered phases to the quantum paramagnetic region, independent of in-plane-angle of the field. 

The field-dependent measurements of specific heat ($C$) and magnetic Gr\"uneisen parameter (\GammaMag{}) 
%\dkc{I made the subscript of $\GammaMag{}$ italic again, as in the PRL.} 
were performed in a dilution refrigerator using the relaxation method for $C$~\cite{Suppl} and the high-resolution alternating-field method for $\GammaMag$~\cite{tokiwa-rsi11}. The background of the cell was subtracted from the raw data, unless stated otherwise~\cite{Suppl}. The orientation of the field along different in-plane directions was achieved by tilting the sample holder with brass wedges of different angles. In this way, we were able to rotate the sample in $5^{\circ}$ steps with the field always pointing in the $ab$-plane~\cite{Suppl}. The same high-quality single crystal as in Ref.~\cite{Bachus2020} was used. It was grown by vacuum sublimation~\cite{Reschke2018}, and its quality was checked by heat-capacity as well as susceptibility measurements~\cite{Bachus2020}.

\begin{figure}[t!]
\includegraphics[width=0.49\textwidth]{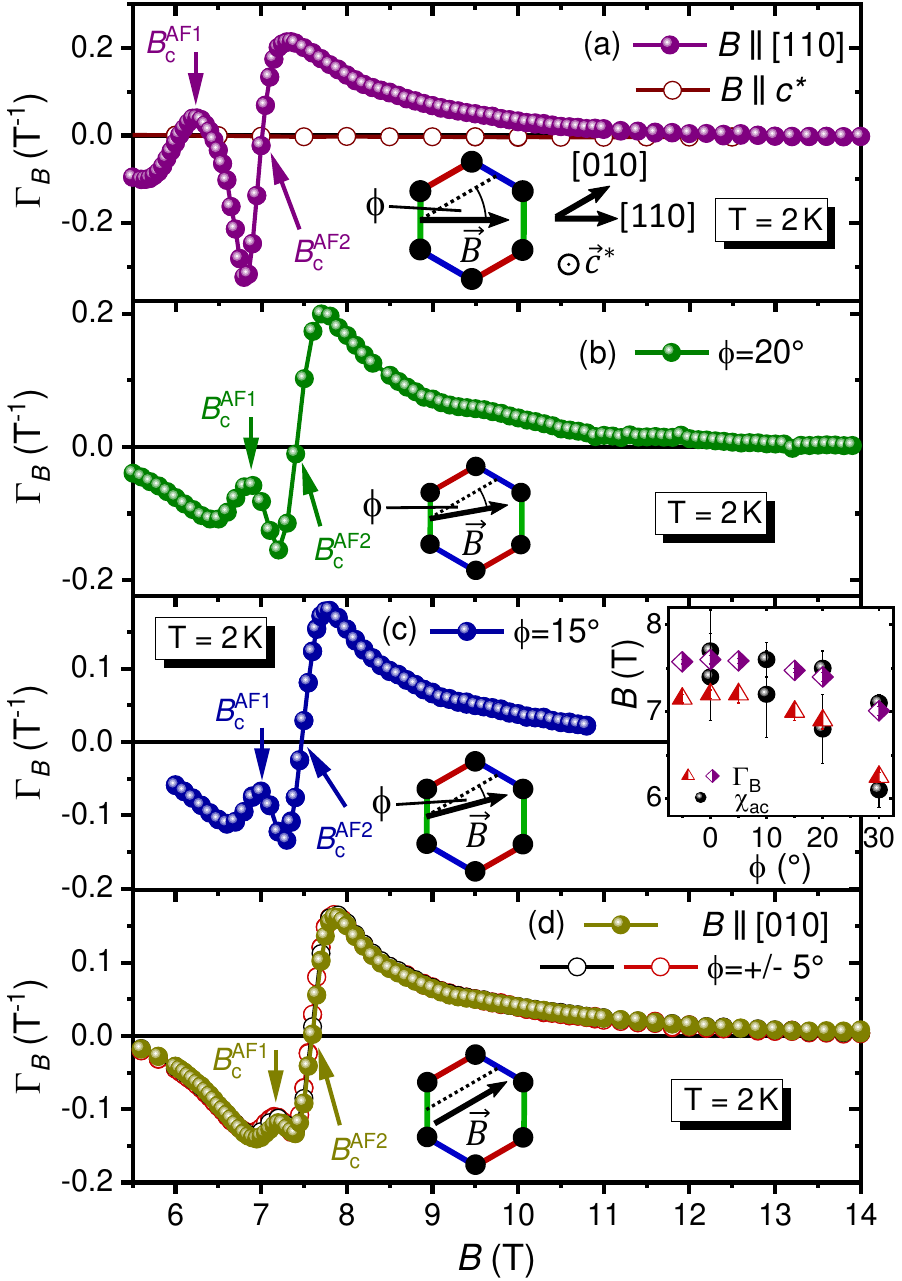}%
	\caption{Angle dependence of the magnetic Gr\"uneisen parameter \GammaMag{} at 2\,K (raw data, see~\cite{Suppl}). The critical fields of the phase transitions \Bc[1] and \Bc[2] are shifted for different in-plane field orientations, from (a) $\vec{B}\parallel\adirection{}$ to (d) $\vec{B}\parallel\bdirection{}$; the data for $\phi=20^{\circ}$ %\dk{Maybe it should be clarified that this could actually be -40 or +80 degree} 
	are from Ref.~\cite{Bachus2020}. The inset in (c) shows the excellent agreement with the critical fields determined from ac-susceptibility measurements~\cite{lampenkelley2018}. No further phase transitions are observed above \Bc[2] for all field directions up to at least 14\,T. The measurement perpendicular to the honeycomb plane ($\vec{B}\parallel \vec{c^*}$) in (a) excludes any out-of-plane contribution.
\label{fig.angleDependentPhaseDiagram}}
\end{figure}%figure* makes graph over whole page

In Fig.~\ref{fig.angleDependentPhaseDiagram}, we show the field dependence of
$\GammaMag (B)=-(\partial M/\partial T)/C$ up to 14\,T for several in-plane
field directions. We also performed a measurement with the field perpendicular
to the $ab$-plane ($\vec{B}\parallel\vec{c^*}$) that returned
$\GammaMag{}\approx 0$ up to 14\,T [empty circles in
Fig.~\ref{fig.angleDependentPhaseDiagram}(a)]. This behaviour is to be expected since no
field-induced transitions should occur in the out-of-plane field within the
field range of our study~\cite{Johnson2015,Majumder2015,Baek2017,riedl2019sawtooth}, and even the N\'eel temperature of $\alpha$-RuCl$_3$
does not change appreciably.
 This confirms that all features observed in our measurements arise from in-plane fields and can not be caused by sample misalignment. 

Our data shows two phase transitions as a function of field strength. The dominant feature is the sign change of \GammaMag{} from negative to positive at \Bc[2], which is equivalent to an entropy maximum at a second-order phase transition. It marks the phase boundary between the long-range-ordered and quantum paramagnetic regions of $\alpha$-RuCl$_3$. A somewhat weaker signature identifies \Bc[1] as a first-order transition between two different antiferromagnetic (AF) ordered states~\cite{lampenkelley2017} caused by a change in out-of-plane ordering~\cite{lampenkelley2018}. In this case, the transition point is determined by a maximum in \GammaMag{}$(B)$, which is equivalent to a smeared negative step in the entropy. Note that Fig.~\ref{fig.angleDependentPhaseDiagram} depicts the raw data of \GammaMag$(B)$ since subtraction of the cell background was not possible for all curves~\cite{Suppl}. This neither affects the values of the critical fields \Bc[1,2], nor changes their evolution at high fields that will be discussed in the following. 

Rotating the field strongly influences the positions of both \Bc[1] and \Bc[2],
illustrating the in-plane anisotropy of the system (compare Fig.~\ref{fig.angleDependentPhaseDiagram} (a)-(d)).
This anisotropy is most
pronounced for the \adirection{}-direction (Fig.~\ref{fig.angleDependentPhaseDiagram}(a)), where already small changes of
$\phi$ influence \Bc[1,2]. Both absolute values of the critical fields and
their angular dependence match perfectly the results of previous
ac-susceptibility ($\chi'$) measurements~\cite{lampenkelley2018}, where
critical fields were determined from peaks in $\chi'$. This validates our
procedure for the evaluation of \Bc[1] and \Bc[2] and allows their measurement
over a broad temperature range, as \GammaMag{} is generally more sensitive to
field-induced phase transitions than, e.g., specific heat.

\begin{figure*}[t!]
\includegraphics[width=0.98\textwidth]{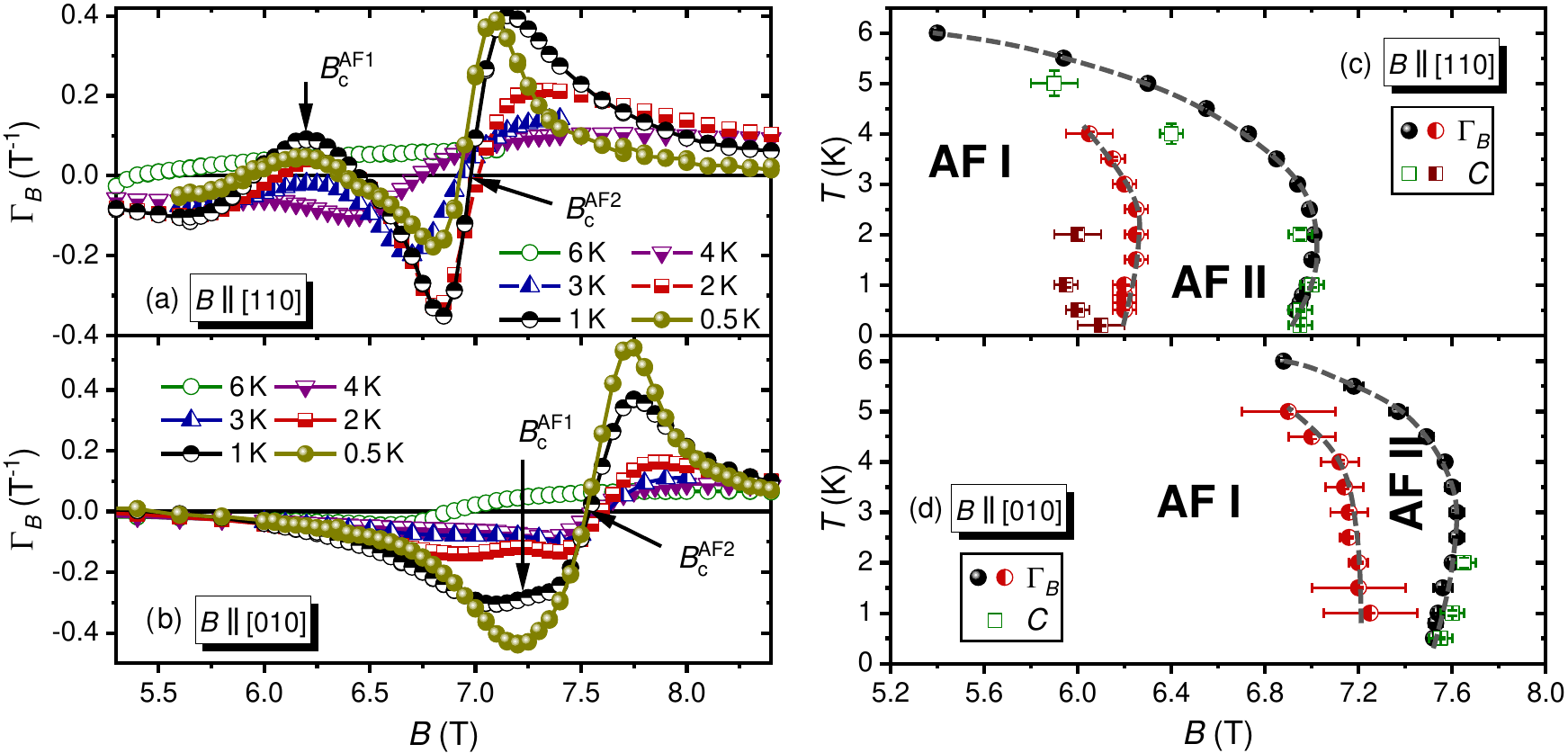}%
	\caption{Field dependence of the magnetic Gr\"uneisen parameter \GammaMag{} (raw data, see~\cite{Suppl}) for several temperatures with the in-plane magnetic field $\vec{B}$ applied parallel (a) to \adirection{}, and (b) to \bdirection{}. \Bc[2] is the point where \GammaMag{} changes sign and indicates the transition between long-range-ordered and quantum paramagnetic states of $\alpha$-RuCl$_3$~\cite{Bachus2020}. At higher temperatures, \Bc[2] shifts towards lower fields due to thermal fluctuations. However, below 2\,K an opposite behavior is observed, suggesting that the phase boundary is non-monotonic (c,d). The phase transition inside the AF state at \Bc[1] is most likely of first-order and, thus, manifests itself in a maximum of \GammaMag{}, which is clearly visible for $\vec{B}\parallel \adirection{}$. By increasing the temperature the maximum is suppressed and gradually shifted toward lower fields. A very similar behavior is observed for $\vec{B}\parallel \bdirection{}$, yet the determination of \Bc[1] is significantly more difficult due to the proximity to \Bc[2]~\cite{Suppl}. The discrepancy between the critical fields extracted from \GammaMag{} and the specific heat $C$ is explained in the main text.
\label{fig.phaseDiagram_a_b}}
\end{figure*}%figure* makes graph over whole page

We will take advantage of this unique sensitivity to probe the presence of field-induced phase transitions above \Bc[2]
%We will take advantage of this unique sensitivity to demonstrate the absence of field-induced phase transitions above \Bc[2] \dk{sounds like we already go into it with this message. we should be neutral observers who come to this finding}, 
but first we discuss peculiarities of the critical fields \Bc[1], \Bc[2] as a function of temperature.

\begin{figure}[t!]
\includegraphics[width=0.49\textwidth]{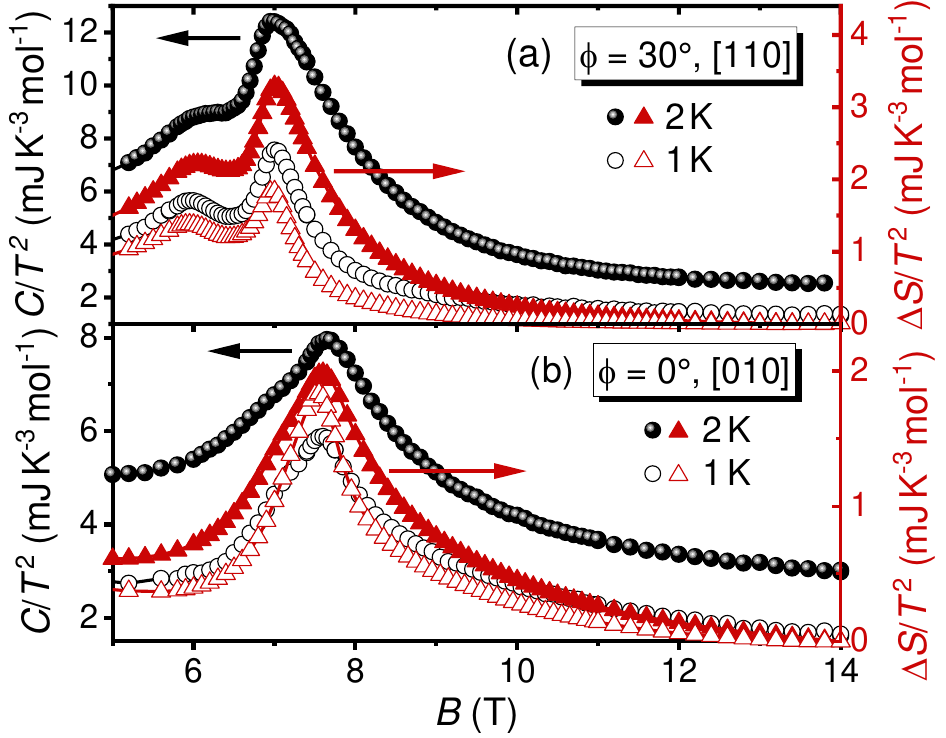}%
	\caption{
%\dk{This is not ``magnetic'' specific heat, but total specific heat, correct? Axis label has $C_m$ - SB: changed, previously it was Cm with phonons subtracted}
%	\dk{Am I the only one who thinks on first sight that these arrows denote ``sweeping direction'' of the experiment, instead of axis label?}
%\dk{It looks like specific heat would be much larger for (b) than for (a), but it is just the y axis scale. }
Field-dependence of specific heat $C$ and entropy increment $\Delta S$ for in-plane fields parallel to (a) \adirection{} and (b) \bdirection{} at 1 and 2\,K, scaled by $T^2$ for better comparison. No signature for any phase transition beyond \Bc[2] is visible up to 14\,T. The most prominent part arises from the peak at the phase transition \Bc[2] for both field directions. While \Bc[1] can also be identified in (a) $C(B)$ and $\Delta S(B)$ by another peak, respectively, it is not pronounced in the measurements for (b), most likely due to the closeness to \Bc[2].
	\label{fig.HC_Entropy_1K2K_a_b}
	%\dk{Ask Sebastian: Can you make a plot of DeltaS as well for $\pm 5^\circ$? Out of interest mainly. }
	}
\end{figure}

Fig.~\ref{fig.phaseDiagram_a_b}(a) shows the field-dependent raw data of \GammaMag{} for several temperatures and the field applied along the \adirection{}-direction. %The field \Bc[2] where \GammaMag{} changes sign, is reduced upon increasing temperature above $\sim 2$\,K.
For $T\geq 2\,$K, the critical field \Bc[2] (marked by the sign change of \GammaMag{}) shifts to lower fields with increasing temperature.  
 This behavior is expected, since the stability of a symmetry-broken phase ($B<\Bc[2]$) is usually decreased by temperature. 
%This anticipated behavior stems from the fact that higher temperatures assist magnetic field in suppressing the long-range order, and the critical field decreases upon heating. 
%Conventionally, one expects the ordered phase to become less stable against the magnetic field with increasing temperature
 However, below 2\,K an opposite behavior is observed, resulting in the non-monotonic phase boundary of \Bc[2]~[Fig.~\ref{fig.phaseDiagram_a_b}c]. A very similar temperature dependence of \Bc[2] is also detected for $\vec{B}\parallel\bdirection{}$ [Fig.~\ref{fig.phaseDiagram_a_b}b,d] and verified by independent specific heat measurements, where \Bc[2] manifests itself by a peak
 [Fig.~\ref{fig.HC_Entropy_1K2K_a_b}]. Below 2\,K, we find excellent agreement between the \Bc[2] values from \GammaMag{} and $C$. At higher temperatures, the peak in the specific heat broadens, and the determination of \Bc[2] from the peak position becomes less accurate than from the sign change in \GammaMag{}. Temperature dependence of \Bc[1] shows an overall similar behavior, except for the fact that the temperature dependence of \Bc[1] is monotonic for $\vec{B}\parallel\bdirection{}$. 

The peculiarities of the phase boundaries are independently confirmed by field
dependence of the magnetic entropy obtained as $\Delta S=-\int dB\,\GammaMag C$
using the Maxwell equation $(\partial S/\partial B)=(\partial M/\partial T)$
[Fig.~\ref{fig.HC_Entropy_1K2K_a_b}]. The Clausius-Clapeyron equation requires
that $d\Bc[1]/dT=-\Delta S/\Delta M$, where $\Delta S$ and $\Delta M$ are
changes in, respectively, entropy and magnetization across the transition.
Since $\Delta M>0$, $d\Bc[1]/dT>0$ implies $\Delta S<0$, and indeed we observe
the reduction in $\Delta S$ around \Bc[1] for $\vec{B}\parallel\adirection$
(Fig.~\ref{fig.HC_Entropy_1K2K_a_b}(a))  but
not for $\vec{B}\parallel\bdirection$ (Fig.~\ref{fig.HC_Entropy_1K2K_a_b}(b)).
%Likewise, for a second-order transition, $d\Bc[2]/dT=-\Delta(dS/dB)/\Delta(dM/dB)$, where $dS/dB=-\GammaMag C$.  \dk{Where do we know from that $\Delta(dM/dB)$ is positive? $\Delta M$ is always positive, but $\Delta(dM/dB)$ is nontrivial (?)} The positive $\Delta(\GammaMag)$ then leads to $d\Bc[2]/dT>0$ in agreement with the positive slope of the phase boundary below 2\,K. 

%%%%%%% INTERPRETATION PHASE BOUNDARY
 
 %Before I undesrtood that Inverse freezing is the scenario that AF2 is a spin glass, not that below AF2 there is a spin glass
%We therefore find a narrow range of magnetic fields, near 7.0\,T for $\vec{B}\parallel\adirection{}$ and 7.6\,T for $\vec{B}\parallel\bdirection{}$, where the system upon cooling first enters a magnetically ordered state and later becomes disordered again. %, suggesting ``inverse melting'' (IM) behavior. 
%If no glassy phase is present, such behavior is known as ``inverse melting'' (IM), or else ``inverse freezing'' (IF). 
%While IF is a well-known phenomenon for spin systems with spin-glass phases \cite{}, we are not aware of an experimental observation of IM for spin systems without quenched disorder \dk{Nature 422, 701–704 (2003) I dont know whether this counts as a spin system. If so, then say ``IM is extraordinary/rare/unique for spin systems so far.}. 

As likely no symmetry is broken for $B>\Bc[2]$, the non-monotonic phase
boundary implies a narrow range of magnetic fields, near 7.0\,T for
$\vec{B}\parallel\adirection{}$ and 7.6\,T for $\vec{B}\parallel\bdirection{}$,
where the system upon cooling first enters a magnetically ordered state (AF2)
and at lower temperatures becomes disordered again. 
%, known as ``inverse melting'' behavior. 
%If no glassy phase is present, such behavior is known as ``inverse melting'' (IM), or else ``inverse freezing'' (IF). 
%While IF is a well-known phenomenon for spin systems with spin-glass phases \cite{}, 
% While we are not aware of experimental obversations of IM in spin systems  \dk{Nature 422, 701–704 (2003): I dont know whether this counts as a spin system. If so, then say ``IM is extraordinary/rare/unique for spin systems so far.},
Such \emph{inverse melting} behavior has been predicted theoretically for some anisotropic spin models~\cite{hui1988reentrant,schupper2005inverse,parente2014reentrant}. %https://reader.elsevier.com/reader/sd/pii/S0304885313009347?token=05DC8E10583C5A43E8DD1C2A2CAE4B9EDBDE98F53DDE99DA819E094ADB7A69C4EF1C4D76029724E40F15B074D1A8DCEC
%this one is scared of finite size effects though
 %Extended Kitaev models may therefore pose an interesting avenue for exploring IM physics. 
 It may therefore be an interesting future avenue for theory to also search for inverse melting in extended Kitaev models.  %Extended Kitaev models should therefore be explored for IM physics. 
  From the integrated entropy differences, we note that the total observed
  field-induced entropy release from zero field, $S(0\,\mathrm T) -
  S(14\,\mathrm T)\approx 0.65\mathrm{mJ}\,\mathrm{mol}^{-1}\,\mathrm K^{-1}$
  (here at $T=1\,K$, $\phi=0^\circ$), appears to be much larger than the
  maximum \textit{magnetic} entropy  that one would expect in the zero-field
  antiferromagnetic phase with a gap $\Delta>1.5\,$meV (inferred from inelastic
  neutron scattering \cite{Ran2017,Banerjee2018}) \footnote{To estimate an \textit{upper} bound of magnetic entropy for a antiferromagnetic phase with gap $\Delta$, one may consider completely flat magnon bands with energy $E(\mathbf q)=\Delta$. For $\Delta=1.5\,$meV and $T=1\,K$ this leads to $S_\mathrm{max}\approx 0.0042\,$mJ\,mol$^{-1}$K$^{-1}$. }. Therefore,
  there are either additional low-energy magnetic excitations at $B=0$ below
  the resolution of neutron scattering \cite{Wellm2018,hentrich2020highfield},
  and/or the significant magnetoelastic coupling of \rucl\
  \cite{Gass2020,kaib2020magnetoelastic,Schoeneman2020,hentrich2020highfield}
  leads to a strongly field-dependent phonon entropy through
  magnetostriction-type effects. 

\begin{figure}[t!]
\includegraphics[width=0.49\textwidth]{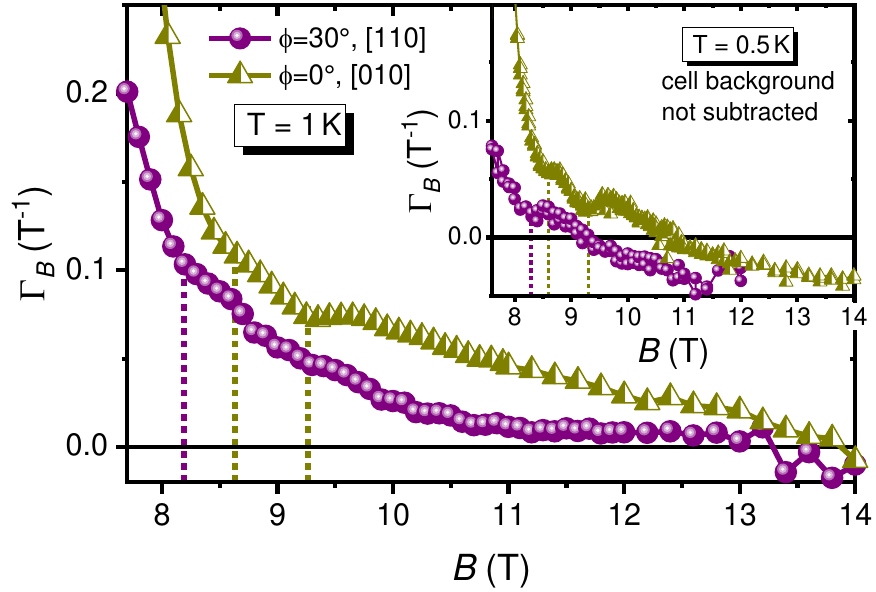}%
        \caption{
%       \dk{I think it looks confusing if the 20$^\circ$ measurements does not have a dotted line for the shoulder?}
Close-up of high-field part of \GammaMag{} at 1\,K. Any purported phase transition is absent above \Bc[2] up to 14\,T, and \GammaMag{} goes to zero due to very tiny remaining entropy [Fig.~\ref{fig.HC_Entropy_1K2K_a_b}]. The shoulder reported in~\cite{Bachus2020} for $\phi =20^{\circ}$ indicative for excited level crossing is also visible for $\vec{B}\parallel \bdirection{}$, roughly setting in at the same position of $\sim 9.3\,$T. Additionally, a weaker second kink is present at $8.6\,$T. For $\vec{B}\parallel \adirection{}$ a similar change of slope can be identified at $\sim 8.2\,$T as the only anomaly above \Bc[2]. Shown in the inset, the measurements at $0.5\,$K confirm the aforementioned results with the lacking phase transition and the more clearly visible anomalies. Note that the negative values of \GammaMag{} are caused by the cell background, which could not be subtracted at $0.5$\,K~\cite{Suppl}.
        \label{fig.Gamma_shoulder_1K} }
\end{figure}

Having established boundaries of the magnetically ordered phases in $\alpha$-RuCl$_3$, we now proceed to the evolution of the system in the quantum paramagnetic region above \Bc[2]. 
%Here, several measurements of thermal Hall effect reported the half-integer plateau at $7-9$\,T~\cite{Kasahara2018} %\dk{actually they applied fields tilted out of the plane and this is the in-plane component.}
Here, several measurements of thermal Hall effect reported the half-integer plateau \cite{Kasahara2018,Yokoi2020,Yamashita2020} --- at $\sim 9$\,T to $\sim 11.5$\,T for $\vec{B}\parallel \adirection{}$ ~\cite{Yokoi2020} %\dkc{I removed the mentions of the plateau-field-strengths for out-of-plane fields. 9.7 to 11.5 is what Yokoi writes himself for $B\parallel a$. However Yokoi has 3 samples with quantization and it sounds like they have slightly different fields}
--- and also suggested an extreme sensitivity of this measurement to the sample quality~\cite{Yamashita2020}. 
If this thermal Hall plateau originates from a gapped topological KSL as suggested by these reports, phase transitions to the bordering topologically trivial phases should occur. 
%One would generally expect that at least one phase transition occurs above
%\Bc[2], should a distinct QSL phase exist in $\alpha$-RuCl$_3$. 
These would happen either through gap closing %(as suggested by numerical studies on the Kitaev model in a magnetic field \cite{})
 or be first order. 
Our data firmly excludes the former possibility for \rucl, since no sign change from negative to positive is observed in \GammaMag{} above \Bc[2] at any field direction [Fig.~\ref{fig.angleDependentPhaseDiagram}] and  temperatures down to 0.5\,K  [Fig.~\ref{fig.Gamma_shoulder_1K}]. 
%for any field direction [Fig.~\ref{fig.angleDependentPhaseDiagram}] and down to down to temperatures of 0.5\,K  [Fig.~\ref{fig.Gamma_shoulder_1K}]. 
%We note that also for a putative gapless KSL 
 Instead, \GammaMag{} appears to asymptotically approach zero at high field strengths for all angles [Fig.~\ref{fig.angleDependentPhaseDiagram}], which would be consistent with the behavior of a single partially-polarized phase above \Bc[2]. 
 Before discussing the alternative scenario of a first-order transition out of a putative KSL, we inspect \GammaMag{} for further possible anomalies aside from sign changes. 
%,which is compatible with a monotonically growing gap in the partially-polarized phase. 

The only conspicuous feature in \GammaMag{} at 1\,K above \Bc[2] is a shoulder
for $\vec{B}\parallel\bdirection{}$, setting in at
$9.3$\,T~[Fig.~\ref{fig.Gamma_shoulder_1K}, yellow symbols]. This can be
paralleled to similar weak anomalies reported previously in the magnetocaloric
effect ($\phi=20^{\circ}$)~\cite{Balz2019} and magnetostriction
(\bdirection{})~\cite{Gass2020}. In contrast, for
$\vec{B}\parallel\adirection{}$, no clear shoulder is visible. However, a
rather abrupt change of slope at $\sim 8.2$\,T is present (Fig.~\ref{fig.Gamma_shoulder_1K}, magenta dotted line), which can be
interpreted as a shoulder superimposed on the rapidly decreasing $\GammaMag{}
\sim 1/(B-\Bc[2])$ in the vicinity of \Bc[2] due to critical
fluctuations~\cite{GarstM:SigctG}. Interestingly, a similar feature can be
identified for \bdirection{}, too [Fig.~\ref{fig.Gamma_shoulder_1K}, yellow dotted
lines]. By reducing the temperature to 0.5\,K, these anomalies become more
prominent, possibly with two shoulders for $\vec{B}\parallel\bdirection{}$
[inset Fig.~\ref{fig.Gamma_shoulder_1K}]. 
%Note, that at 0.5\,K no background subtraction was possible which explains the
%negative \GammaMag{}~\cite{Suppl}.
Previously, we assigned such a shoulder feature %for $\phi=20^\circ$
 to a level crossing in low-energy excitations~\cite{Bachus2020}. 
%Our data in Fig.~\ref{fig.Gamma_shoulder_1K} suggest a universal nature \dk{I would tone this down. Suggests that this is the case for all shoulder-anomalies in other systems} of such a level crossing that happens around the same field regardless of its in-plane direction. 
Our data in Fig.~\ref{fig.Gamma_shoulder_1K} suggest that the field strength at which this occurs has only a weak dependence on the in-plane field direction, qualitatively consistent with an \textit{ab-initio}-derived microscopic  model of \rucl~\cite{kaib2020magnetoelastic,Suppl}. 
This observation is also in line with recent spectroscopic measurements suggesting only a weak dependence of $\Gamma$-point
($\mathbf q=0$) excitations on the direction of the in-plane field~\cite{Wulferding2020,Ponomaryov2020}. %For \bdirection{}, however, an additional crossing seems to appear. 

%To investigate the possibility whether these anomalies or other features in our data could could support a first-order transition out of a KSL, we examine field-dependent entropy changes, obtained as $\Delta S=-\int dB\,\GammaMag C$. These are shown for $\phi=0^\circ$ and $\phi=30^\circ$ in Fig.~\ref{fig.HC_Entropy_1K2K_a_b}, to be read with the right axis. 

 While we ruled out a continuous transition above \Bc[2], we now investigate the possibility whether these anomalies and/or other features in our data could support a first-order transition out of a  KSL phase. Such a first-order scenario was put forward recently \cite{Tanaka2020}, 
where both the putative gapless ($\phi=0^\circ$) and gapped ($\phi\neq 0^\circ$) KSLs would %have first-order phase boundaries towards a gapped topologically trivial phase 
experience first-order transitions towards a gapped topologically trivial phase at $B \gtrsim 10\,$T. 
  We therefore examine field-dependent entropy changes, obtained as $\Delta S=-\int dB\,\GammaMag C$. %These are shown in Fig.~\ref{fig.HC_Entropy_1K2K_a_b} for $\phi=30^\circ$ (where the putative KSL would have the largest gap) and for $\phi=0^\circ$ (putative gapless). %to be read with the right axis. 
  These are shown for $\phi=30^\circ$ and $\phi=0^\circ$  in Fig.~\ref{fig.HC_Entropy_1K2K_a_b}, where each entropy curve has been shifted such that $\Delta S(14$\,T)$=0$\,mJ\,K$^{-1}$\,mol$^{-1}$.  
%  Each entropy curve has been shifted so that $\Delta S(14$\,T)$=0$\,mJ\,K$^{-1}$\,mol$^{-1}$.  
  %Our $\Delta S$ at 1\,K and 2\,K do not show signatures of latent heat freed at such a transition.
No anomalies are detected in $\Delta S$ above \Bc[2] for either field direction and up to 14\,T. 
%While a first-order transition between two gapped states with low entropy as proposed for $\phi=30^\circ$ may only show little signatures here, a putative gapless KSL 
For $\phi=30^\circ$, the gap within the putative KSL would be the largest, such that a first-order transition to another phase with comparable gap may be hard to detect. 
However, for the putative gapless KSL along $\phi=0^\circ$, one would expect significant latent heat to be released at a first-order transition to a gapped state, which we however do not observe. 
%   Especially for a putative gapless KSL ($\phi=0^\circ$), significant latent heat should be released at a first-order transition to a gapped state, which is however not observed. 
%      Especially for a putative gapless KSL ($\phi=0^\circ$), significant latent heat should be released at a first-order transition to a gapped state, which is however not observed. 
   Instead, entropy appears to shrink asymptotically with field strength, consistent with a continuously growing gap in a single partially-polarized phase above \Bc[2].

In summary, our comprehensive thermodynamic study of $\alpha$-RuCl$_3$ revealed several universal features of this material that do not depend on the direction of the in-plane magnetic field. For the magnetically ordered phases $B\leq \Bc[2]$, the phase boundary separating them from the quantum paramagnetic state is non-monotonic, suggesting a narrow region of inverse melting. The in-plane anisotropy manifests itself in the different stability range of the intermediate ordered phase observed between \Bc[1] and \Bc[2].
For the quantum paramagnetic region $B>\Bc[2]$, our data is clearly inconsistent with the existence of an additional continuous transition and also speaks against a first-order transition. This applies both for field angles where a gapless Kitaev spin liquid (KSL) and angles where a gapped KSL have been proposed, implying that the plateau in the thermal Hall conductivity in \rucl\ does not go along with a topologically nontrivial KSL. 
%Second, no additional phase transitions are observed beyond this phase boundary. This leaves no room for Kitaev QSL as a distinct magnetic phase of $\alpha$-RuCl$_3$ \dk{there are some asteriks... Phase transition could be way above 14\,T. Here one may have to argue more (for example: spectroscopic measurements undoubtedly observe $B\gg 10$\,T or $B\gtrsim 14$ as the partially-polarized phase)}. Third, magnetic Gr\"uneisen parameter \GammaMag{} is positive above \Bc[2] suggesting a steady decrease in entropy upon increasing magnetic field at a constant temperature, and consistent with the presence of an excitation gap. 
 %, and in the dependence of the excitation gap on the field direction. 
 Instead, the observed thermodynamics are qualitatively consistent with a single phase above $\Bc[2]$ with a monotonically growing excitation gap. 
%\dk{The observed inverse melting presents an interesting challenge for future theoreitcal work on extended kitaev models}

%------------------------------------------------------------------------------------------------------

\begin{acknowledgments}
We thank Yoshi Tokiwa for his experimental assistance during early stages of this work. The research in Augsburg was supported by the German Research Foundation (DFG) via the Project No. 107745057 (TRR80) and by the Federal Ministry for Education and Research through the Sofja Kovalevskaya Award of Alexander von Humboldt Foundation (AAT). The work in Frankfurt was supported by the DFG Project No. 411289067 (VA117/15-1). RV  was supported in part by the National Science Foundation under Grant No. NSF PHY-1748958.
\end{acknowledgments}

%---------------------------------------------------------------------------------

%\bibliography{RuCl3_angleDependence}

\clearpage

\begin{widetext}
\begin{center}
\large{\textbf{\textit{Supplemental Material}\smallskip\\
Angle-dependent thermodynamics of $\alpha$-RuCl$_3$}}
\end{center}
%\tableofcontents

\renewcommand{\thefigure}{S\arabic{figure}}
\renewcommand{\thetable}{S\arabic{table}}
\renewcommand{\theequation}{S\arabic{equation}}
\setcounter{figure}{0}
\setcounter{equation}{0}

\section{Variation of in-plane field direction}
The magnetic field orientation in the honeycomb plane of \aRuCl{} was adjusted by tilting the cell, including the sample platform, using brass wedges with different angles [Fig.~\ref{figSM.variation_in_plane}(a,b)]. The crystal was mounted with the honeycomb planes on the platform and the cell is attached to the cryostat sample holder, such that the magnetic field always points in the planes. Rotating the cell by the wedges results in a variation of the in-plane field angle. This way, we measured several configurations and determined the critical fields \Bc[1,2]. By comparing these results to ac-susceptibility measurements~\citep{lampenkelley2018s}, we calibrated our field orientation relative to the [010] direction which we defined as $\phi =0^{\circ}$. Here, we used the near six-fold symmetry which has further been validated by the angle-dependency of specific heat~\cite{Tanaka2020s}. Note that our previously measured field direction $\phi=20 ^{\circ}$~\cite{Bachus2020s} as defined in the main text used the setup described in Fig.~\ref{figSM.variation_in_plane}c. Strictly speaking, this corresponds to a value of $\phi= 100 ^{\circ}$, which is equivalent to $\phi = 20^{\circ}$ in the angle-field phase diagram~[Fig.~1, main text]. In the same way, $\phi = -5^{\circ}$ is equivalent to $\phi = +5^{\circ}$. The excellent agreement of our angle-field phase diagram with the ac-susceptibiliy measurements further approves the six-fold symmetry of the \Bc[1] and \Bc[2] transitions.
\begin{figure*}[b!]
\includegraphics[width=0.98\textwidth]{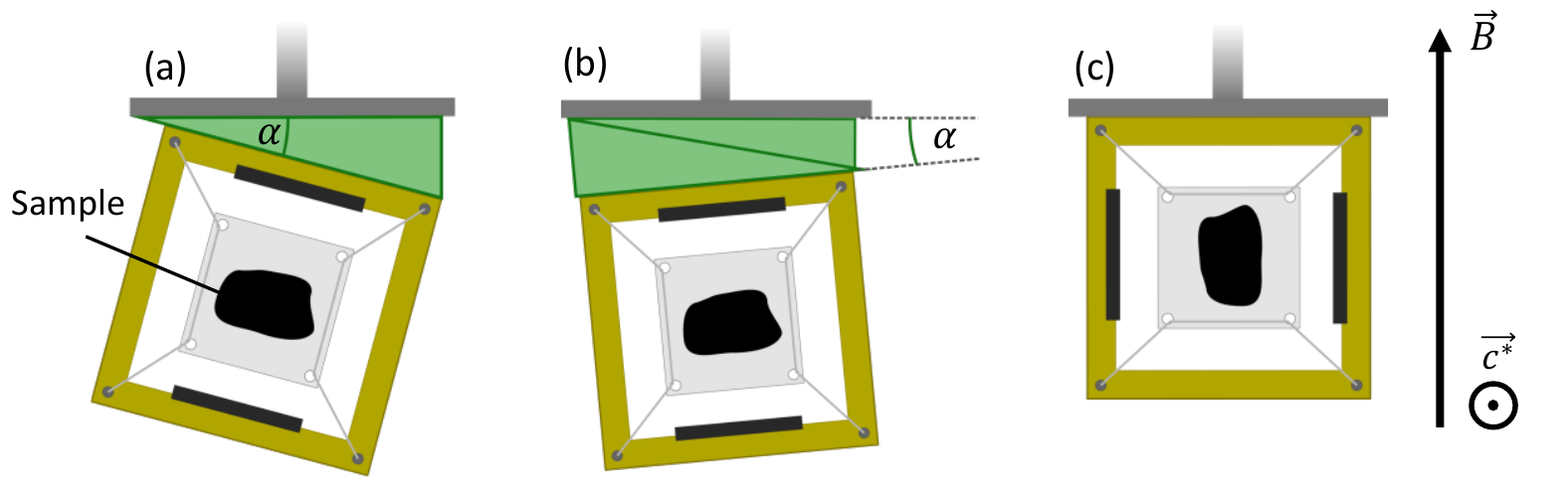}%
	\caption{(a) Rotation of Gr\"uneisen and specific heat setup by using brass wedges with the angles $\alpha =10/ 15/ 20^{\circ}$. The sample is mounted with the honeycomb planes being parallel to the sample platform ($\vec{c^*}$ perpendicular to sample platform, magnetic field orientation indicated in the sketch). (b) Angles of $\pm 5/25/30^{\circ}$ can be achieved by combining two wedges. In this way, rotating the cell by $\pm 30^{\circ}$ in $5^{\circ}$ steps is realizable. Consequently, the magnetic field is orientated in different directions in the honeycomb planes. Note, however, that $\alpha$ is not equivalent to $\phi$ from the main text. We used the wedges to align the sample with several in-plane field directions. Afterwards, we determined the respective field orientation $\phi$ relative to the [010] direction which we calibrated using the critical fields \Bc[1,2] from ac-susceptibility~\cite{lampenkelley2018s}, see Fig.~1 in the main text, together with the near six-fold symmetry~\cite{lampenkelley2018s,Tanaka2020s}. The in-plane field direction $\phi = 20^{\circ}$ as defined in the main text was measured previously~\cite{Bachus2020s} with the configuration shown in (c). Strictly speaking, this corresponds to a value of $\phi= 100 ^{\circ}$ which is equivalent to $\phi = 20^{\circ}$ in the angle-field phase diagram~[Fig.~1, main text]. In the same way, $\phi = -5^{\circ}$ is equivalent to $\phi = +5^{\circ}$.
	\label{figSM.variation_in_plane}}
\end{figure*}

%\section{Qualitative results for C(B) below 1\,K}

\section{Cell background contribution for magnetic Gr\"uneisen parameter \GammaMag{} below 1\,K}
\begin{figure*}[b!]
\includegraphics[width=0.98\textwidth]{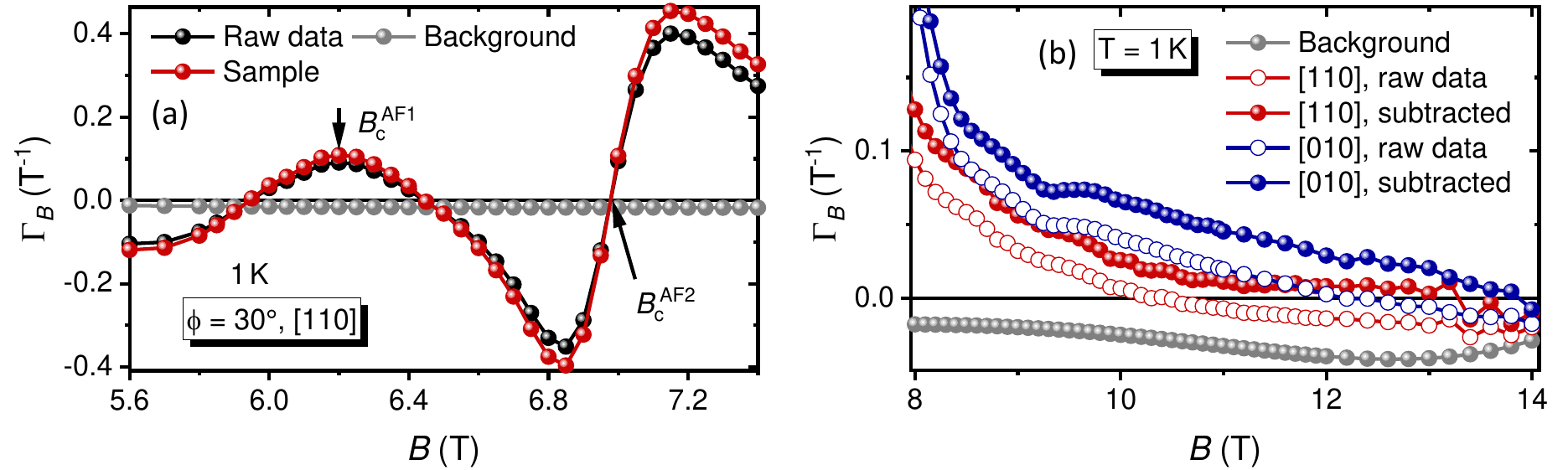}%
	\caption{(a) Cell background subtraction for [110] at 1\,K following Eq.~\eqref{eq.GammaH_Background}. Raw data denote the Gr\"uneisen parameter from both cell and sample. For the background measurement, \GammaMag[,Cell] was measured independently. Despite the subtraction, the positions of \Bc[1] and \Bc[2] are unaffected. This fully justifies using the raw data for constructing the phase diagram in the main text. (b) Influence of the cell background at high fields. Here, the sample exhibits a specific heat $C$ and Gr\"uneisen parameter \GammaMag{} comparable to the background (see also~\cite{Bachus2020s}). As a consequence from the negative background, the raw data cross zero at higher fields, too (open circles). However, after subtraction, the Gr\"uneisen parameter \GammaMag[,Sa] of the sample remains positive (full circles).
	\label{figSM.Gamma_background_subtraction}}
\end{figure*}
%\dk{I advocate for making clear all the time that background subtraction refers to the cell background, not the intrinsic phonon background of RuCl. So I added ``cell'' to ``background'' in many instances. }
Samples with a small heat capacity compared to that of the cell require a careful background subtraction for the Gr\"uneisen parameter \GammaMag{}. First, the total value \GammaMag[,tot] has to be measured which includes both cell and sample contributions. Second, the heat capacities \Csample{} and \Ccell{} of, respectively, the sample and cell are needed. Together with \GammaMag[,Cell], the sample's Gr\"uneisen parameter is calculated according to Ref.~\cite{Bachus2020s}:
\begin{equation}
\GammaMag[,Sa]=\GammaMag[,tot]+\frac{\Ccell}{\Csample}\left(\GammaMag[,tot]-\GammaMag[,Cell]\right).
\label{eq.GammaH_Background}
\end{equation}
As a result, the background subtraction is not feasible in the absence of $C(B)$. Considering huge time requirements for such a measurement, as opposed to the measurement of \GammaMag{}, we chose not to perform it for each temperature and field direction and used the raw data in our analysis. 

In Fig.~\ref{figSM.Gamma_background_subtraction}, we show exemplary background subtraction for \GammaMag{} at 1\,K. One sees that the cell background is negligible in the vicinity of \Bc[1] and \Bc[2], so the raw data can be safely used for the determination of critical fields. The background contribution becomes more significant above \Bc[2], where specific heat of the sample decreases. Here, the background is responsible for the apparent negative values of \GammaMag{} above $10-12$\,T, while positive values are recovered once the background contribution is subtracted. Therefore, we argue that \GammaMag{} of the sample remains positive even well above \Bc[2]. We also note that, should any phase transitions occur above \Bc[2], they would of course show up also in the raw data.

%measurement or valid analysis is available, a background subtraction for \GammaMag{} is not possible. Since specific heat measurements take significantly longer than measuring the magnetocaloric effect used for determining \GammaMag{}, several $\GammaMag(B)$ sweeps do not obtain the specific heat counterpart due to time considerations. Consequently, in these cases, a background subtraction was not feasible and the raw data are shown, e.g., in Fig.~1 in the main text. However, the position of \Bc[1] and \Bc[2] is not affected. In Fig.~\ref{figSM.Gamma_background_subtraction}a, the field dependency for $B \parallel$[110] at 1\,K is depicted. Both raw and subtracted data exhibit the same values for \Bc[1] and \Bc[2]. Accordingly, the temperature-field phase diagram is not affected by a lacking background subtraction. 

%The same holds true for fields above \Bc[2] where specific heat and Gr\"uneisen parameter decrease quickly. Any clear signature comparable to \Bc[1,2] would be visible in the raw data as well. Therefore, the raw data of \GammaMag{} at 500\,mK [inset Fig.~3, main text] exclude any phase transition beyond \Bc[2]. The only consequence of the background are the apparently negative values at higher fields. Yet, this is not resulting from the sample signal. As shown in Fig.~\ref{figSM.Gamma_background_subtraction}b for two distinct field orientations at 1\,K, the cell background is obviously responsible for this behavior. After the subtraction, \GammaMag[,Sa] diminishes rapidly while staying positive.

\section{More details about \GammaMag{}}
\begin{figure*}[t!]
\includegraphics[width=0.98\textwidth]{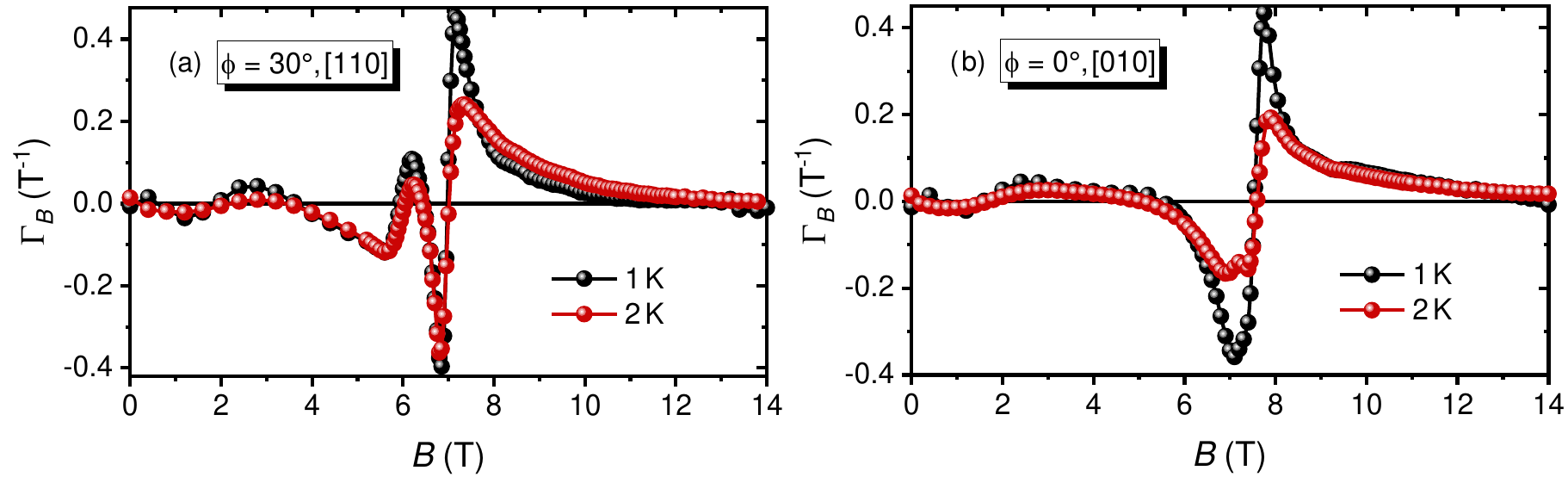}%
	\caption{Field dependence of the magnetic Gr\"uneisen parameter \GammaMag{} over the whole measured range from 0 to 14\,T for the different field orientations (a) [110] and (b) [010], respectively. The main signatures are the transitions at \Bc[2] and \Bc[1], the latter being weakest for the [010] direction. The sign change at $\sim$2\,T from~\cite{Bachus2020s} is reproduced and likely results from domain reconstruction~\cite{Sears2017s}.
	\label{figSM.Gamma_whole_field_range}}
\end{figure*}
\begin{figure*}[ht]
\includegraphics[width=0.98\textwidth]{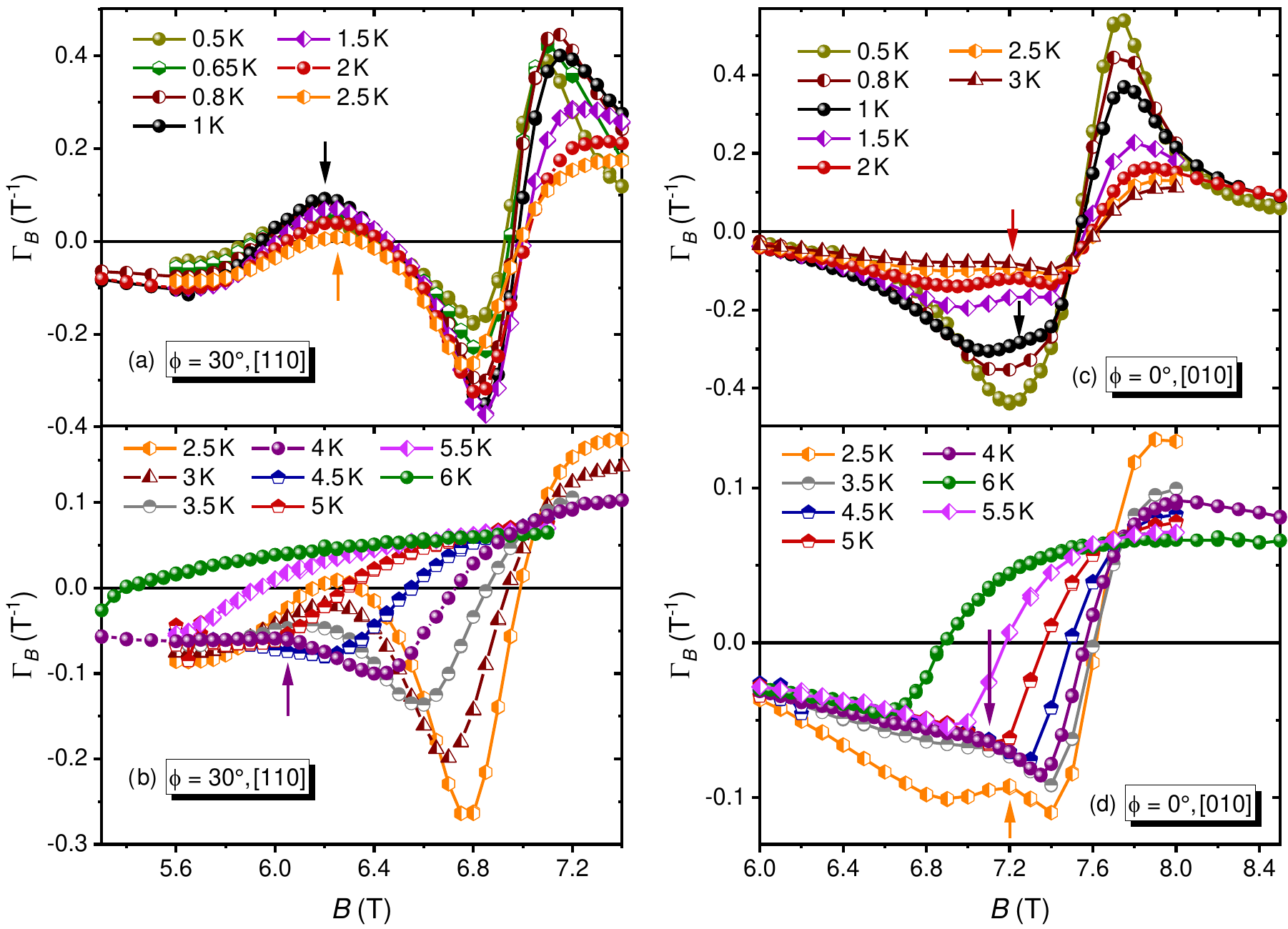}%
	\caption{Detailed \GammaMag{} measurements used for the determination of \Bc[1] and \Bc[2] for (a,b) [110] and (c,d) [010]. Raw data without cell background subtraction are shown, because corresponding specific heat data were not always available. This does not affect the positions of \Bc[1] and \Bc[2], as explained in the text. The transition \Bc[2] is easily identified as the sign change from negative to positive for all temperatures and field directions. (a,b) [110] direction. The maximum indicative of \Bc[1] is clearly visible at temperatures below 2.5\,K (black and orange arrow) and broadens toward higher temperatures transforming into a kink at 4\,K (purple arrow), and finally vanishes. (c,d) [010] direction. Here, the determination of \Bc[1] is hindered by the proximity to the dominant transition at \Bc[2]. The 2\,K data show a maximum (red arrow) that develops into a broad shoulder at 1\,K (black arrow), and becomes fully screened by the increasingly sharp feature at \Bc[2] toward lower temperatures. Above 2.5\,K, the maximum becomes a kink (orange and purple arrow) and vanishes, similarly to the [110] orientation.
	\label{figSM.Gamma_determination_Bc1}}
\end{figure*}
\begin{figure*}[t!]
\includegraphics[width=0.98\textwidth]{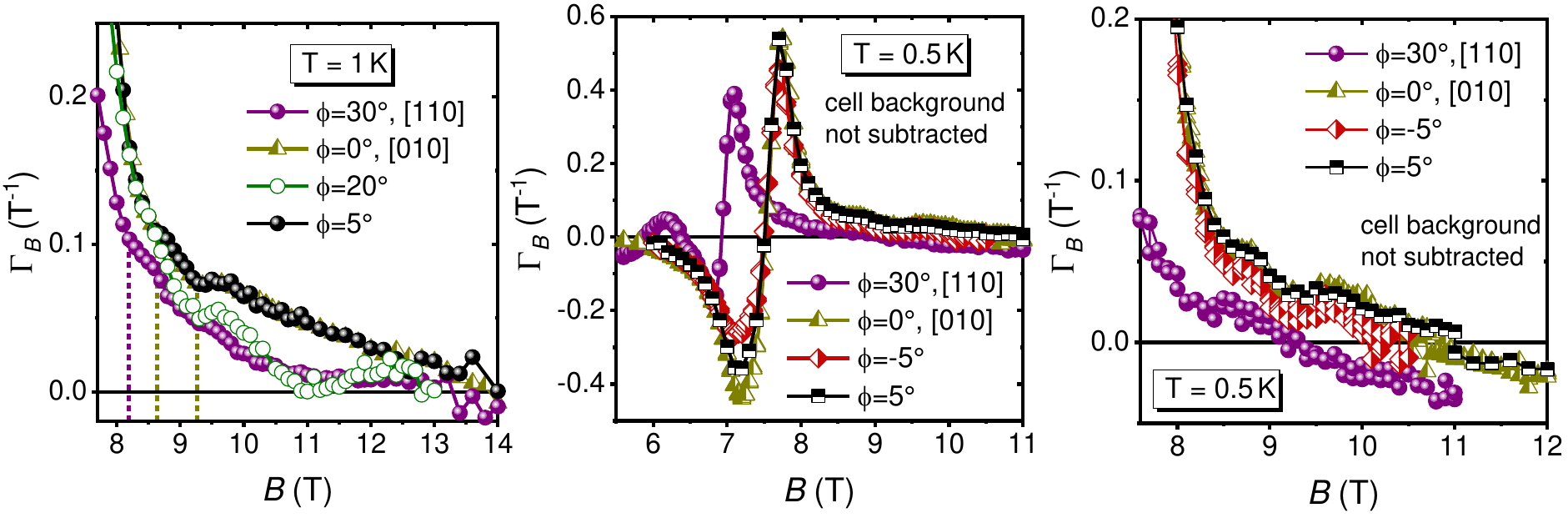}%
	\caption{
	%Field dependence of \GammaMag{} at 500\,mK for different field direction. Here the raw data are shown due to the lack of $C(B)$ data like explained previously. 
Field dependence of \GammaMag{} at (a) 1\,K and (b,c) 500\,mK (no background subtraction available due to lack of $C(B)$). (a) More field directions at 1\,K. (b) At 500\,mK, \Bc[1] and \Bc[2] are very well resolved for all field directions. As already shown previously, the subtraction of the background does not change the position of any anomaly in $\GammaMag{}(B)$. (c) Zoom-in into \GammaMag{} above \Bc[2]. Shoulder-like anomaly(ies?) visible which is most pronounced for $\vec{B}\parallel \vec{b}$.
	\label{figSM.Gamma_500mK}}
\end{figure*}
Here, we present our Gr\"uneisen parameter study in more detail. Fig.~\ref{figSM.Gamma_whole_field_range} illustrates the whole field range from 0 to 14\,T for the field applied along (a) [110] and (b) [010] at 1 and 2\,K, respectively. As already discussed before, the main signatures result from \Bc[2] and \Bc[1], the latter being weakest for [010]. The sign change due to an entropy maximum at $\sim$2\,T from~\cite{Bachus2020s} is reproduced and likely related to domain reconstruction, as reported previously~\cite{Sears2017s}.

Now, we focus on establishing the temperature-field phase diagram from the main text [Fig.~2]. In Fig.~\ref{figSM.Gamma_determination_Bc1}, all measured temperatures are shown for (a,b) [110] and (c,d) [010], respectively. Determining \Bc[2] as transition from the AF to the field-polarized state is straightforward due to the obvious sign change from negative to positive. In contrast, \Bc[1] can not always be identified with comparable accuracy. For [110] at temperatures up to 2.5\,K, the maximum indicative of the first-order transition is clearly visible [arrows in Fig.~\ref{figSM.Gamma_determination_Bc1}(a)]. However, this maximum broadens towards higher temperatures and eventually becomes almost completely smeared out at 4\,K. At this point, \Bc[1] is defined by the remaining kink [arrow in Fig.~\ref{figSM.Gamma_determination_Bc1}(b)]. Above that temperature, even this kink fades away. With the field oriented along [010], establishing \Bc[1] is even more challenging because of the proximity to the dominant feature at \Bc[2]. At 1\,K, \Bc[1] appears as a broad shoulder, which evolves into a maximum at higher temperatures, e.g.\ 2\,K [arrows in Fig.~\ref{figSM.Gamma_determination_Bc1}(c)]. Yet, it can not be identified any more below 1\,K. Similarly to [110], the maximum transforms into a kink at higher temperatures [Fig.~\ref{figSM.Gamma_determination_Bc1}(d)] and eventually vanishes.

Finally, the field-evolution of the Gr\"uneisen parameter at 1\,K and 500\,mK is shown in Fig.~\ref{figSM.Gamma_500mK} for several field orientations. At 1\,K in Fig.~\ref{figSM.Gamma_500mK}(a), $\phi =20^{\circ}$ reveals a plateau at similar position like [010]. Furthermore, the weaker anisotropy along [010] is corroborated because rotating by $\phi =5^{\circ}$ does not change $\GammaMag{}(B)$ signicifcantly compared to [010]. At 500\,mK, the cell background can not be subtracted as explained above. A minimal difference between $\phi =\pm 5^{\circ}$ directly below \Bc[2] might indicate that $\phi =0^{\circ}$ is not perfectly aligned along [010]. Nonetheless, this misalignment should be smaller than $2^{\circ}$. Aside from this small deviation, the data for $\pm 5$ and $0^{\circ}$ are almost perfectly on top of each other, further confirming the weaker anisotropy along the [010] direction. Fig.~\ref{figSM.Gamma_500mK}(c) represents the evolution above \Bc[2] in more detail. As described in the main text, the shoulder feature as a universal characteristic is visible for all field orientations. The measurements of $\phi =\pm 5^{\circ}$ confirm the probable appearance of a second shoulder anomaly for $\phi = 0^{\circ}$.

\section{Results of heat capacity below 1\,K}
\begin{figure*}%[t!]
\includegraphics[width=0.98\textwidth]{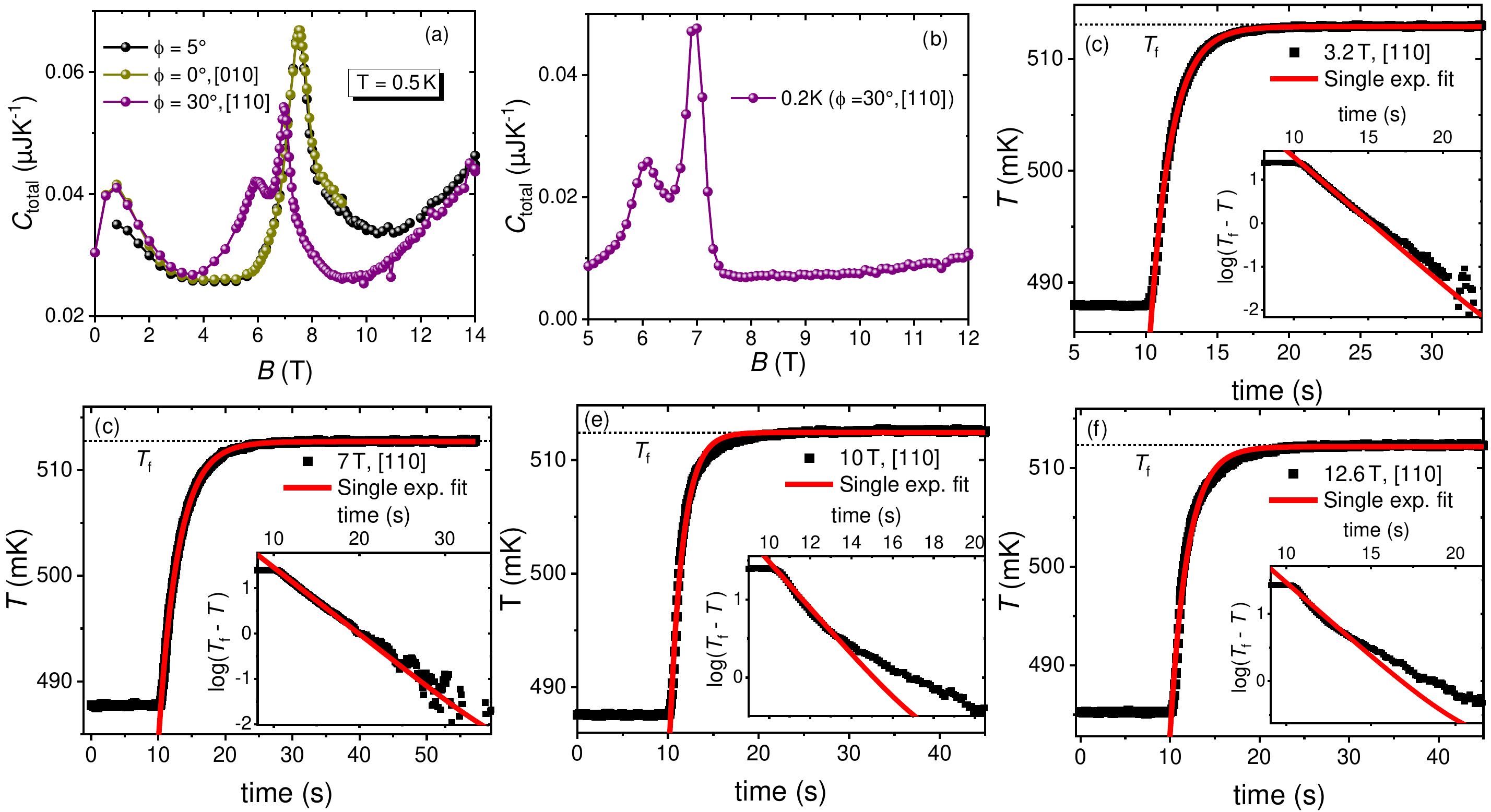}%
	\caption{(a) Field dependence of the total heat capacity \Ctot{} (including background) up to 14\,T for several field orientations. They perfectly match the 1\,K data from the main text, undoubtedly resolving \Bc[2] and, in the case [110], also \Bc[1]. Far above \Bc[2], fitting with a single exponential function is not adequate anymore resulting in the unphysical increase towards higher fields due to shortcomes of the single exponential function analysis. Nevertheless, we speculate that a phase transition beyond \Bc[2] would give rise to a deviation of the monotonic behavior, e.g., in form of another peak. As a consequence, from these data, we suspect that a phase transition seems to be unlikely. (f) Heat capacity for [110] at 200\,mK. \Bc[1] and \Bc[2] are perfectly resolved while the monotonic behavior beyond \Bc[2] does not indicate any further phase transition. (c-f) Examples for the sample temperature response \temp{} at 500\,mK and different magnetic fields, all applied along the [110] direction. A single exponential function describes \temp{} very well for lower fields of e.g.\ (a) 3.2\,T and (b) 7\,T, respectively. At higher fields well beyond the phase transition \Bc[2], the single exponential fit obviously deviates from the data points at (c) 10\,T and (d) 12.6\,T, respectively. This is even more obvious in the logarithmic plot in the insets where obviously two time constants are present, manifesting themselves in different slopes.
	\label{figSM.HC_analysis_500_200mK}}
\end{figure*}
In this part, we present field-dependent heat capacity measurements below 1\,K. First, we focus on the phase transitions \Bc[1] and \Bc[2]. Second, we discuss the influence of nuclear contribution on the results above \Bc[2]. Finally, we conclude with a comment about the analysis at low temperatures and high fields in our setup, and how this influences a potential determination of a small gap. 

Fig.~\ref{figSM.HC_analysis_500_200mK}(a) shows the field-dependent heat capacity of the \aRuCl{} crystal at 500\,mK for several directions in absolute units of \textmu{}J/K since no cell background subtraction was possible, as explained in more detail below. The phase transition at \Bc[2]$\sim 7-7.6$\,T is perfectly resolved as a peak for all field orientations. For [110], a second maximum indicative for \Bc[1] is clearly visible and used as well for the phase diagram in the main text. In a similar manner, the heat capacity data at 200\,mK and [110] orientation are interpreted [Fig.~\ref{figSM.HC_analysis_500_200mK}(b)]. Both \Bc[1] and \Bc[2] are undoubtfully identifiable.

Now, we return to the 500\,mK data [Fig.~\ref{figSM.HC_analysis_500_200mK}(a)].
%Towards higher fields a monotonic increase is observed being unreasonable for \aRuCl{} and discussed in more detail below. Nevertheless, any further peak comparable to the clear \Bc[2] transition is lacking since no deviation of the monotonic behavior is detectable for any field direction. The same holds true at 200\,mK and the field along [110]. Again, no evidence for strong signatures analogous to \Bc[2] is present. This way, we speculate that our heat capacity data show the absence ...
Towards higher fields a strange monotonic increase is observed, its origin will be discussed below. Notwithstanding this spurious effect, no additional anomaly can be seen above \Bc[2] for any field direction, also for the 200\,mK data in the field along [110]. Therefore, we conclude that our heat capacity data show the absence of further transitions beyond \Bc[2] for all measured field directions.  

In the following, we focus on the reason for the unphysical heat capacity behavior far above \Bc[2]. We attribute this to the presence of nuclear contribution in our setup. This contribution is enhanced in high magnetic fields and becomes comparable to the small sample heat capacity in the order of several $\sim 10$\,nJ/K. Combined with a sizable spin-lattice relaxation time this may give rise to the so-called 2$\tau$-effect~\cite{Andraka1996,Andraka2011_note} and causing discrepancies when standard analysis with a single exponential function is used. Obviously, as shown in Fig.~\ref{figSM.HC_analysis_500_200mK}(c-f), such deviations only occur at fields far beyond \Bc[2], which is unambiguously visible in the logarithmic plots (insets). Consequently, the analysis using the single exponential function results in unreasonable behavior. We assign the nuclear contribution to the sapphire platform (Al$_2$O$_3$) due to the nuclear moments of Al because the background measurement without sample revealed very similar behavior. Accordingly, a precise analysis of the cell background was impossible, too, preventing a background subtraction in Fig.~\ref{figSM.HC_analysis_500_200mK}(a,b).

As a result, due to this problem below 1\,K and at high fields, our heat capacity data do not allow to determine the excitation gap under high magnetic fields. The framework beyond Refs.~\cite{Andraka1996, Andraka2011_note} would be needed to ensure an appropriate subtraction of the nuclear contribution and goes beyond the scope of our present work.
%are not feasible for determining the field gap. This would require highly accurate measurements in exactly that problematic temperature and field regime because temperatures significantly below the potential gap would be essential. However, our measurement setup and the resulting heat flow scheme is different compared to the situation in~\cite{Andraka1996, Andraka2011_note}, and, thus, a varied fit function would be needed for a more suitable 2tau analysis, going beyond the scope of this work.

\section{Comparison to extended Kitaev models}
To compare our measurements of \GammaMag{} to expectations from such extended Kitaev models, we employ exact diagonalization on a two-dimensional 24-sites honeycomb cluster with $C_3$ symmetry. To access \GammaMag{} at small finite temperatures, we follow the method applied for the theoretical calculations in Ref.~\cite{Bachus2020s}.  While these two-dimensional calculations cannot capture the transition \Bc[1] related to the change of the out-of-plane ordering wave vector, other properties like \Bc[2], the magnitude and sign of \GammaMag{} and possible occurrences of shoulder anomalies may be compared qualitatively. 

In extended Kitaev models, the three nearest-neighbor bonds of the honeycomb lattice are defined as X$_1$, Y$_1$, Z$_1$ bonds depending on their orientation. Second-neighbor X$_2$, Y$_2$, Z$_2$ (third-neighbor X$_3$, Y$_3$, Z$_3$) bonds within the plane are then defined as those orthogonal (parallel) to the directions of the respective X$_1$, Y$_1$, Z$_1$ bonds. 
The effective magnetic Hamiltonian for Z$_n$-bonds ($n\in\{1,2,3\}$) then reads
\begin{equation}
	H_\mathrm{Z_n} = \sum_{\langle ij\rangle_{\mathrm Z_n}}  J_n\, \mathbf S_i \cdot  \mathbf S_j + K_n \, S_i^z S_j^z + \Gamma_n\, \left(S_i^x S_j^y + S_i^y S_j^x\right) + \Gamma'_n\, \left(S_i^x S_j^z + S_i^z S_j^x + S_i^y S_j^z + S_i^z S_j^y\right),
	\label{eq:zbond}
\end{equation}
with effective spin-$\frac12$ operators $\mathbf S_i$. The Hamiltonians for X$_n$ and Y$_n$ bonds can be obtained by cyclic permutation of spin components $(x,y,z)$ in Eq.~\eqref{eq:zbond}. Furthermore, a Zeeman term $H_\text{Zee}=-\mu_B\sum_i \mathbf B \cdot \mathbf g\cdot \mathbf S_i$ can contribute with magnetic field $\mathbf B$ and the g-tensor $\mathbf g$. 
If $K_1$ is the only finite exchange coupling and $B=0$, the model reduces to the exactly solvable Kitaev honeycomb model. 
\begin{figure*}
\includegraphics[width=0.75\textwidth]{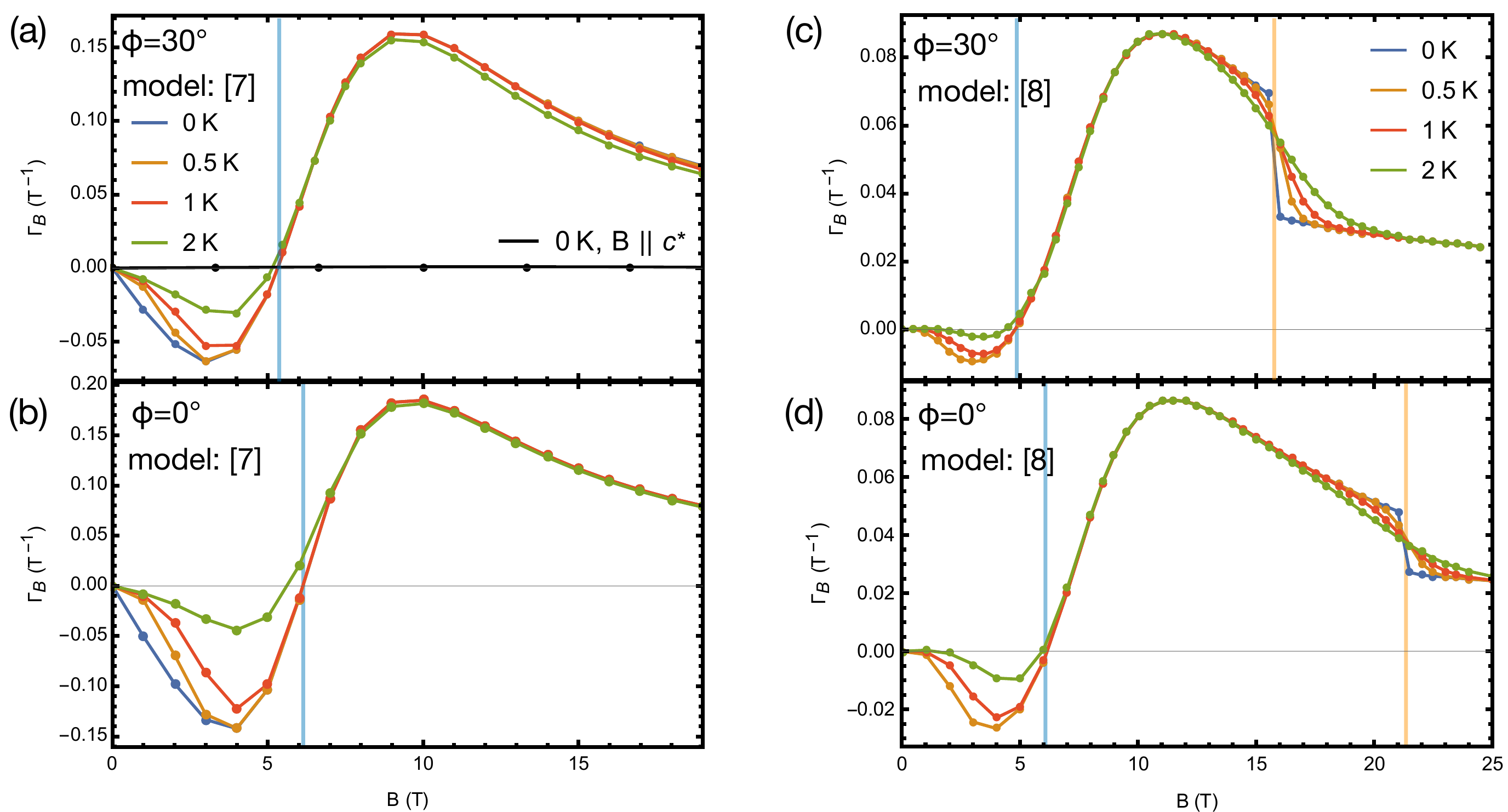}%
	\caption{Numerical results for \GammaMag{} in two extended Kitaev models and different field directions. 
	Blue vertical lines mark sign changes in \GammaMag{}, orange lines mark shoulder anomalies. 
	(a,b) Results for the model of Ref.~\cite{Winter2018}. (c,d) Results for the model of  Ref.~\cite{kaib2020magnetoelastics}, whose small second-neighbor $J_2$ and Dzyaloshinskii-Moriya interactions were neglected. 
	\label{fig:theory}
	}
\end{figure*}

We discuss two representative models that have been shown to be qualitatively consistent with numerous experimental observations in $\alpha$-RuCl$_3$. 
%In both these cases, no in-plane-field-induced Kitaev spin liquid is found for in-plane magnetic fields after the suppresion of zigzag order. 
%Neither model is found to feature a field-induced Kitaev spin liquid after the suppression of zigzag order, and so, far up to our knowledge, no model has been found that does so. 
Fig.~\ref{fig:theory}(a,b) shows results for \GammaMag{} for the model of Ref.~\cite{Winter2018}, whose nonzero magnetic couplings are $(J_1,K_1,\Gamma_1,J_3)=(-0.5,-5,2.5,0.5)\text{\,meV}$ and where $(g_{ab},g_{c^\ast})=(2.3,1.3)$.  Fig.~\ref{fig:theory}(c,d) shows results for the \textit{ab-initio} parameters of  Ref.~\cite{kaib2020magnetoelastics}, $(J_1, K_1, \Gamma_1, \Gamma'_1, K_2, \Gamma_2, \Gamma'_2, J_3, K_3, \Gamma_3, \Gamma'_3)=(-5.66, -10.12, 9.35,-0.73,-0.18,0.06,0.03,0.2,0.25,0.04,-0.07) \text{\,meV}$ and $g_{ab}=2.36$, where we omitted weak second-neighbor $J_2$ and Dzyaloshinskii-Moriya exchanges. 
Both models feature zigzag antiferromagnetic order at zero magnetic field and a single phase transition under in-plane magnetic fields to a partially-polarized phase. Within the antiferromagnetic phase, \GammaMag{} is found to be negative up to the critical fields (corresponding to the suppression of magnetic order, \Bc[2]) near to 6~Tesla. 
At these critical fields, a sign change in \GammaMag{} occurs. The fact that \GammaMag{} does not diverge at \Bc[2] and only shows a delayed maximum at higher field strengths is likely a finite-size effect of the calculation, as the gap can not fully close at a continuous transition on a finite cluster. 
Various universal features found in our measurements are nevertheless captured qualitatively: The field strength at which the sign change takes  place shows the correct dependence on the in-plane angle in both models [blue vertical lines in Fig.~\ref{fig:theory}], with highest magnetic fields for $\phi=0^\circ$ and lowest for $\phi=30^\circ$. While the model of Ref.~\cite{Winter2018} shows no anomalies beyond the sign change and maximum in $\GammaMag{}$, the model of Ref.~\cite{kaib2020magnetoelastics} features shoulder-anomalies at higher field strengths [orange vertical lines in Fig.~\ref{fig:theory}(c,d)]. Such anomalies are found in various extended Kitaev models that are proximate to competing phases at finite fields, but do not enter them \cite{Bachus2020s}. They occur as a result of field-induced level crossings in the lowest excited states within the partially-polarized phase and appear as a negative jump in $\GammaMag{}$ for $T\rightarrow 0$\,K, that is smeared out at finite temperatures. In the present model of Ref.~\cite{kaib2020magnetoelastics}, both the position of the sign change, and that of the should-anomaly are found to shift to lower field strengths upon rotating from $\phi=0^\circ$ to $\phi=30^\circ$. This behavior is qualitatively consistent with the anomalies observed in experiment [Fig.~4 of main text], however in the present set of \textit{ab-initio} parameters occurs at much higher field strengths. This might be a result of finite-size effects and/or could be refined by adjusting these parameters. 
We also checked the effect of magnetic fields perpendicular to the honeycomb plane ($\mathbf B \parallel c^\ast$) [black points in Fig.~\ref{fig:theory}(a)]. In accordance with our measurement, the Gr\"uneisen parameter is found to be approximately zero for this direction (up to field strengths of $\sim 33$\,T in the model of Ref.~\cite{Winter2018}).

%In Fig.~\ref{fig:theory}, we show results for two models, namely that of Ref. \cite{Winter2018,kaib2020magnetoelastic}

\end{widetext}

\end{document}